\documentclass[lettersize,journal]{IEEEtran}
\usepackage{amsmath,amsfonts}
\usepackage{algorithmic}
\usepackage{algorithm}
\usepackage{array}
\usepackage[caption=false,font=normalsize,labelfont=sf,textfont=sf]{subfig}
\usepackage{textcomp}
\usepackage{stfloats}
\usepackage{url}
\usepackage{verbatim}
\usepackage{graphicx}
\usepackage{cite}
\usepackage{makecell}
\usepackage[dvipsnames]{xcolor}

\newcommand{\ignore}[1]{}

\begin{document}

\title{NeuralEQ: Neural-Network-Based Equalizer \\for High-Speed Wireline Communication}

\author{Hanseok Kim,~\IEEEmembership{Graduate Student Member,~IEEE,}
        Jae Hyung Ju,
        Hyun Seok Choi,\\
        Hyeri~Roh,~\IEEEmembership{Graduate Student Member,~IEEE,}
        and~Woo-Seok~Choi,~\IEEEmembership{Member,~IEEE}
\thanks{This work was supported by Samsung Electronics Co., Ltd (Contract ID: MEM210728\_0001).}
\thanks{H. Kim, J. H. Ju, H. S. Choi, H. Roh, and W.-S. Choi are with the Department of Electrical and Computer Engineering and the Inter-University Semiconductor Research Center, Seoul National University, Seoul 08826, South Korea (e-mail: anjeo@snu.ac.kr, wooseokchoi@snu.ac.kr).}%
\thanks{H. Kim is also with Samsung Electronics, South Korea.}%
}



\maketitle

\begin{abstract}
With the growing demand for high-bandwidth applications like video streaming and cloud services, the data transfer rates required for wireline communication keeps increasing, making the channel loss a major obstacle in achieving low bit error rate (BER).
Equalization techniques such as feed-forward equalizer (FFE) and decision feedback equalizer (DFE) are commonly used to compensate for channel loss in wireline communication, but they have limitations in terms of noise boosting and timing constraints. 
On the other hand, the forward-backward algorithm can achieve better BER performance, but its high complexity makes it impractical for wireline communication. 
In this work, we propose a novel neural network, NeuralEQ, that effectively mimics the forward-backward algorithm and performs better than FFE and DFE while reducing complexity of the forward-backward algorithm. 
Performance of NeuralEQ is verified through simulations using real channels.

\end{abstract}

\section{Introduction}
\label{sec:intro}

\IEEEPARstart{R}{ecent} advances in machine learning, video streaming, and cloud services have positively impacted our daily lives by providing us with new and improved ways to access and consume information, entertainment, and services. However, all of these technologies rely heavily on significant amounts of data computation and transportation. 
The I/O bandwidth and computing power of high-performance computing systems must be scaled up accordingly to support large data transactions between computing cores. 
Otherwise, the limited data transfer rates will degrade the system performance.
To meet such ever-increasing bandwidth demand for high-performance computing systems, 
the PCIe standard is being developed up to 64\,Gbps for the next generation, 
and data-rate of NVlink specializing in GPU communication reaches 50\,Gbps~\cite{9547041}. 
Data-rate of the ethernet protocol is also being developed up to 112\,Gbps to handle the high network load.

Although demands for wireline communication with higher data-rate keep increasing,
limited bandwidth of wireline channels (e.g. PCB channels, coaxial cables) poses problems in high-speed data transmission. 
Pursuing low latency and good I/O energy efficiency,
wireline communication has relied on simple modulation like pulse-amplitude modulation with two levels (PAM2) and simple equalizers, followed by symbol-by-symbol detection. 
However, data-rate is increasing so rapidly that the transmitted signal is severely distorted even in short channels, which makes it difficult to restore data at the receiver.

To mitigate this issue, instead of transmitting binary baseband signals, researchers have begun to exploit the multicarrier modulation scheme like orthogonal frequency division multiplexing (OFDM) for wireline communication~\cite{dmt0,dmt1,dmt2}.
This approach utilizes different frequency bands to transmit data simultaneously, allowing data transfer with higher spectral efficiency over lossy channels.
However, adopting multicarrier modulation in wireline communication is not compatible with traditional wireline communication systems, and it can be expensive to update the hardware of network switches in data centers and network infrastructure to support this new approach. 

A rather simpler approach to increasing data transfer rates using baseband signal transmission is to exploit multi-level signaling such as PAM4, which encodes multiple bits of data into a single symbol. 
Since this allows for a higher data-rate than traditional binary signaling,
many applications~\cite{9516687,9830299,9366063} have started to consider adopting PAM4 signaling.
In addition, digital signal processing (DSP)-based equalizers and analog-to-digital converters (ADCs) on the receiver side can also be used to enhance the performance and efficiency of data transfer. 
These technologies can help improve the signal-to-noise ratio (SNR) and reduce interference, allowing faster and more reliable data transfer.
While PAM4 signaling enables data-rate over 100\,Gbps~\cite{9459179,9134399,8952650}, it comes with some challenges. 
One of the main issues is that due to the peak power constraint, PAM4 suffers from low SNR and makes the signal more vulnerable to inter-symbol interference (ISI) caused by limited channel bandwidth. 
This leads to degradation in bit error rate (BER) performance. 
Moreover, since PAM4 signaling requires more advanced receiver designs and equalization techniques,  this can increase the complexity of the communication system.
Under these circumstances, equalizers to compensate for the channel loss play a critical role in high-speed wireline communication.

Equalizers can be primarily divided into linear equalizers (e.g. feed-forward equalizer (FFE)) and nonlinear equalizers (e.g. decision-feedback equalizer (DFE)). 
FFE is a discrete-time finite-impulse-response (FIR) filter that boosts the input in the frequency range where channel loss is high.
It is simple to implement, but since the incoming noise is also amplified, symbol detection after FFE shows limited BER performance. 
On the other hand, DFE does not boost noise and shows better BER, but the timing constraint of the feedback loop makes it difficult to design for high speeds. 
The loop-unrolling, or speculative, technique mitigates the timing constraint but increases hardware complexity significantly, especially in PAM4, resulting in increased power overhead. 
Moreover, DFE alone cannot remove the pre-cursor ISI.
Therefore, high-speed wireline communication systems typically use both FFE and DFE to improve BER~\cite{8534279}.

Recently, different types of equalizers have been also studied. 
Thanks to the development of deep learning, 
receivers using neural networks, especially recurrent neural networks (RNNs)~\cite{Ye,ZHOU2019121,Kechriotis,Kim,Gomez}, 
showed excellent performance for sequence detection. 
However, for high-speed wireline communication, RNNs have the disadvantages that high-speed implementation is difficult due to the timing constraint, i.e. past computation results are required for the current operation. 

Instead of RNNs, fully-connected layers (FCLs) could be used for channel equalization in wireline communication systems.
FCLs are a type of the neural network layer that connects every input neuron to every output neuron, allowing the network to learn complex non-linear relationship between inputs and outputs. 
However, FCLs also have disadvantages that they can be computationally inefficient because they are typically over-parameterized. 
Hence, in some cases, a more sparse connection such as convolutional layers may be more appropriate and efficient.

In view of these, in this paper, a novel neural network architecture is proposed that addresses the limitations of conventional equalizers such as FFE and DFE in high-speed wireline communication systems. 
Compared to RNN-based equalizers, the proposed architecture uses a feed-forward architecture without any data-dependency-induced timing constraints, allowing an efficient pipelined hardware architecture for high-speed receivers. 
Additionally, it has less computational complexity than FCL-based equalizers by removing unnecessary connections between layers. 
This makes the proposed neural network a suitable solution for high-speed wireline communication where both computational efficiency and hardware design constraints are essential factors to consider.
The design is inspired by the forward-backward (FB) algorithm, which is widely used in Hidden Markov Models (HMMs) to estimate the posterior probabilities of the HMM states.



This paper makes the following contributions:
\begin {itemize}
\item Inspired by the FB algorithm, a novel neural network called NeuralEQ is proposed, which has better BER performance than conventional equalizers and lower computational complexity than the FB algorithm.
\item NeuralEQ is designed to be more efficient compared to prior neural network based equalizers in terms of computational complexity and hardware design constraints, making it a suitable solution for high-speed wireline communication.
\item NeuralEQ is found to be highly prunable, where more than half of its weights can be removed without affecting performance. Additionally, NeuralEQ is robust against ISI variation.
\end{itemize}
\section{Background}
\label{sec:background}

The core idea of this paper is 
to develop a novel neural network architecture that mimics the FB algorithm in order to improve the performance of existing equalizers such as FFE and DFE.
Hence, first we introduce the FB algorithm and its properties, then describe how the FB algorithm is mapped to the proposed NeuralEQ.

\begin{figure*}[!t]
\centering
\subfloat[]{\includegraphics[width=0.6\textwidth]{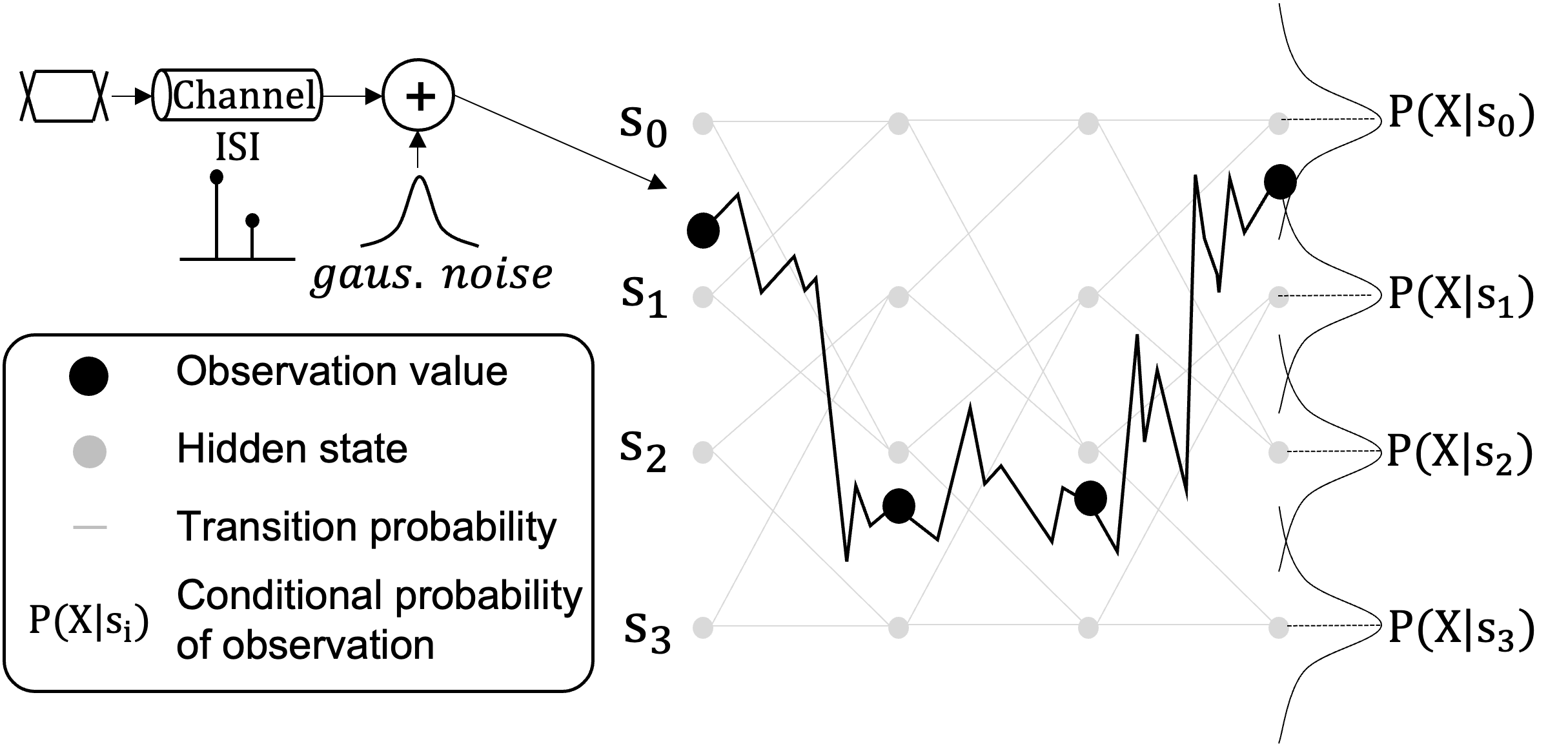}%
\label{fig:wc_hm}}
\subfloat[]{\includegraphics[width=0.4\textwidth]{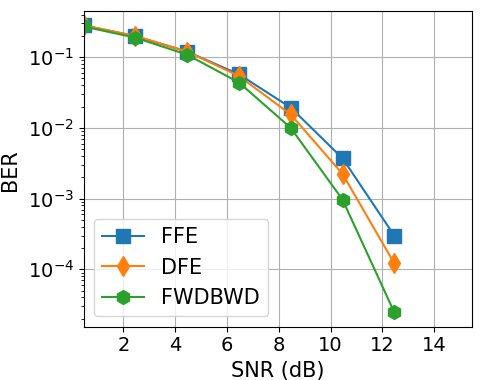}%
\label{fig:isi0_woNEQ}}
\caption{(a) Applying HMM to wireline communication, and (b) performance comparison between FFE, DFE, and FB.}
\end{figure*}

\subsection{Forward-Backward (FB) Algorithm}
\label{sec:fb}

The FB algorithm, which is also known as BCJR~\cite{1055186} or maximum a posteriori (MAP) decoder, is widely used in modern wireless communication with its excellent performance. 
However, since wireline communication requires a much higher data-rate and low latency, running a computationally heavy algorithm such as FB is challenging. 
Described in detail below is the FB algorithm to understand its computational complexity.

The FB algorithm infers the posterior probability of all the hidden states in an HMM.
Let $S^{t}$ be a hidden state random variable at a time $t$, whose value is one of the elements in a set of hidden states $\mathcal{S}$, 
and $X^{1:t}$ be an observed random sequence from time $1$ to $t$. 
If we denote the transition probability from $j$-th state to $i$-th state as $a_{ji}$ in the HMM, then the FB algorithm is summarized as follows. 

The forward probability $\alpha_i^t$ for $i \in \mathcal{S}$ and $1 \leq t \leq T$ is defined as $P(S^{t}=i,X^{1:t}=x^{1:t})$, which can be calculated by the following recursion: 
\begin{equation} \label{equa:fwdbwd}
\alpha _{i}^{t} = \sum _{j\in \mathcal{S}} \alpha _{j}^{t-1}a_{ji}P(X^{t}=x^{t}|S^{t}=i)
\end{equation}
The backward probability $\beta_i^t$ is defined as $P(X^{t+1:T}=x^{t+1:T}|S^{t}=i)$, and the update rule is. 
\begin{equation} \label{equa:bwd}
\beta _{i}^{t-1} = \sum _{j\in \mathcal{S}} \beta _{j}^{t}a_{ij}P(X^{t}=x^{t}|S^{t}=j)
\end{equation}
From $\alpha_i^t$ and $\beta_i^t$, the posterior probability of the state being $i$ at time $t$ for a given sequence from 1 to $T$, i.e. $P(S^t=i|X^{1:T}=x^{1:T})$, defined as $\gamma_i^{t}$, can be calculated as following. 
\begin{equation} \label{equa:gamma}
\gamma_i^{t} =  \frac{\alpha_i^t\beta_i^t}{\sum_{j \in \mathcal{S}} \alpha_j^t\beta_j^t}
\end{equation}
Since the denominator is common for all $i$, MAP estimate of the state at a given time is given by:
\begin{equation}
\hat{S}^{t|T, MAP} =  \arg\max_{i \in \mathcal{S}}{\gamma_i^t} = \arg\max_{i \in \mathcal{S}}{\alpha_i^t\beta_i^t}
\end{equation}

\subsection{Applying Forward-Backward Algorithm to Wireline Communication with Lossy Channel}

The FB algorithm can be applied by modeling wireline communication with a lossy channel as HMM. 
In the following, for this modeling, the channel input is assumed to be a random sequence because typically wireline links are tested with random data patterns.
Note that, although transmitted data are random, due to the channel ISI, each symbol at the channel output is affected by the adjacent symbols.
Specifically, if the discrete-time impulse response of a channel is $h[n]$, the noiseless channel output sequence is computed as the convolution between the input sequence and $h[n]$.

In oreder to apply FB, four parameters in HMM should be defined:
(1) the observed sequence ($X^{1:T}$), (2) the hidden states ($\mathcal{S}$), (3) the state transition probability ($a_{ij}$, for $i, j \in \mathcal{S}$), and (4) the conditional probability of observables given each hidden state ($P(X^t|S^t)$). 
First, the observed sequence is defined as the channel output distorted by ISI and noise. 
Second, each hidden state in $\mathcal{S}$ is defined as the noiseless channel output, which is determined by the channel input sequence with the length of $h[n]$, or $|h[n]|$. 
For example, in PAM4 signaling, there are $4^{|h[n]|}$ possible hidden states, and the set of hidden states is $\mathcal{S} = \{\sum_{i=1}^{|h[n]|}x[i]h[i] \mid x[i] \in \{-1, -1/3, 1/3, 1\}\}$. 
Third, each hidden state has four possible state transitions, but two of them have zero probability and the other two have the equal probability 1/2. 
Finally, assuming an additive white Gaussian noise channel with the noise variance of $\sigma^2$, the conditional probability density of the observable can be written as:
\begin{equation} \label{equa:cond}
\begin{aligned}
P(X^t=x|S^t=s_i)=\frac{1}{\sqrt{2\pi\sigma^2}}\exp[-\frac{1}{2}(\frac{x-s_i}{\sigma})^2]. 
\end{aligned}
\end{equation}
As an example, Fig.~\ref{fig:wc_hm} shows how PAM2 communication through a channel with one-tap post-cursor can be modeled with HMM with the described four parameters.

In order to compare the performance of the FB algorithm with the conventional equalizers,
the following simulation is conducted.
PAM4 data are transmitted through a channel with the impulse response $h[n]=[1.0, 0.4, 0.2, 0.1]$ (3 post-cursors), 
and the FB algorithm is applied after modeling the system with HMM.
For comparison, FFE and DFE taps are chosen as 8 and 3, respectively, which are sufficient to compensate for ISI. 
As demonstrated in Fig.~\ref{fig:isi0_woNEQ}, exploiting the FB algorithm in wireline communication results in much better BER performance compared to the conventional equalizers such as FFE and DFE. 
However, the cost is its computational complexity. 
As described in Section~\ref{sec:fb}, 
the number of hidden states increases exponentially with the length of ISI, 
and the forward and backward computations increase proportionally to the number of hidden states. 
There have been efforts to reduce the complexity in wireless communication~\cite{1603379,6030705,4201022}, 
but it is still challenging to use it in high-speed wireline communication at the data-rate over 100\,Gbps,
and even if implemented, power consumption will be prohibitive.
In view of these drawbacks,  
a neural-network-based equalizer superior to conventional equalizers in terms of BER performance, while reducing the complexity of the FB algorithm, is developed in the next section.

\begin{figure}[t]
\centering
\includegraphics[width=\columnwidth]{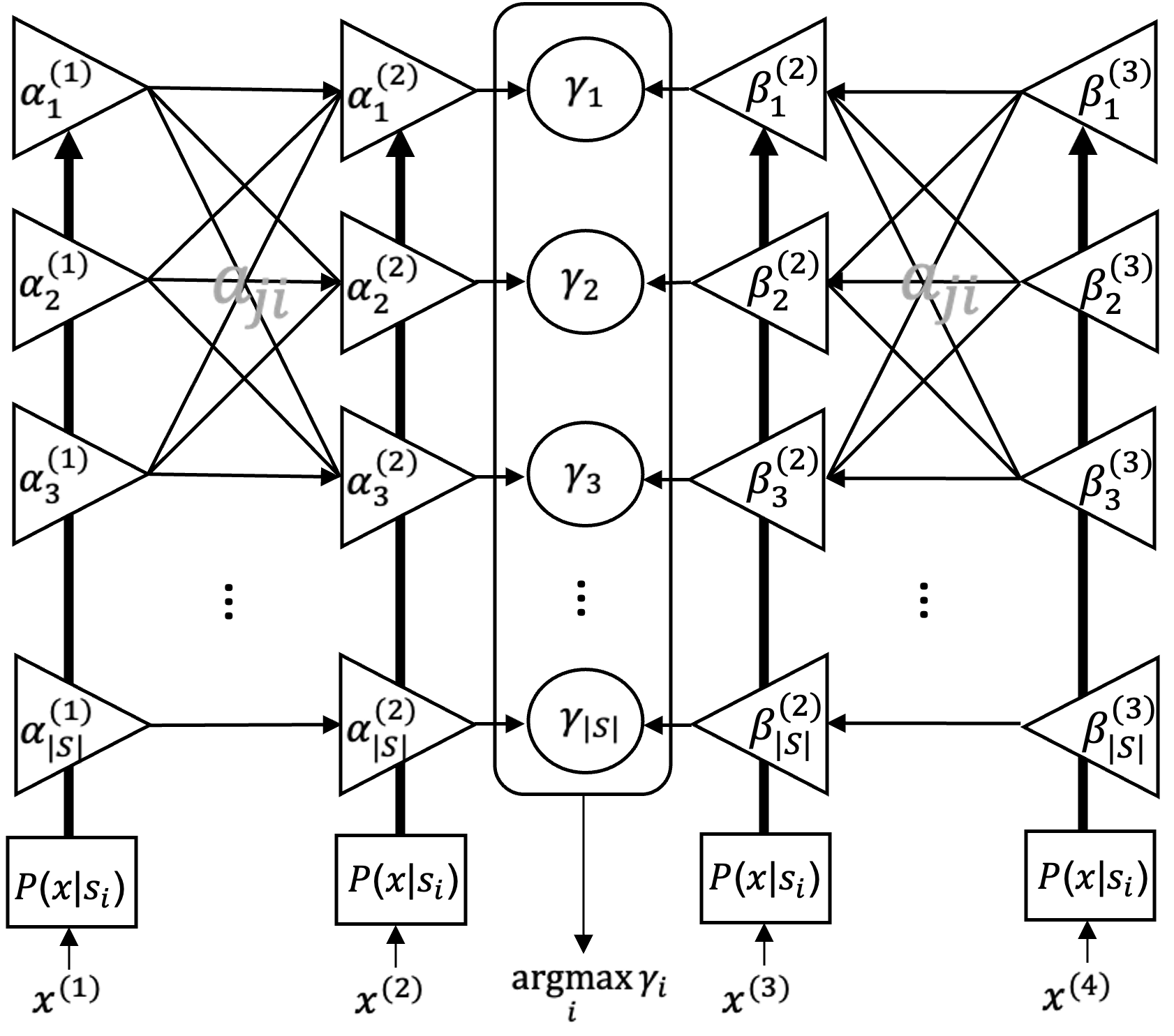}  
\caption{Computational graph of FB algorithm.}
\label{fig:blockFwdBwd}
\end{figure}

\section{Proposed NeuralEQ}
\label{sec:proposed}

There would be various ways to create an equalizer using the neural network. 
The first method would be to use the FCLs~\cite{9079933,8054694,app9214675}.
The fully connected structure has the advantage of being able to mimic any function, 
but it often has unnecessarily many connections to implement the target function. 
There is another way to utilize RNNs. 
Many recent works such as \cite{Ye,ZHOU2019121,Kechriotis,Kim,Gomez} implement equalizers using RNNs because they have excellent performance as a sequence detector, and a receiver can be seen as a kind of sequence detector.
However, because RNN has a recurrent structure that requires computation results from the past inputs in the current operation, 
it is difficult to satisfy the timing margin when designing hardware requiring a high-speed operation.
In view of these drawbacks, we propose a novel neural network architecture that is more suitable for high speed wireline communication than FCL and RNN-based equalizers. 
In this section, we will describe our proposed architecture and demonstrate its effectiveness against FCL and RNN-based equalizers.

\begin{figure*}[ht]
\centering
\subfloat[]{
  \includegraphics[width=0.25\textwidth]{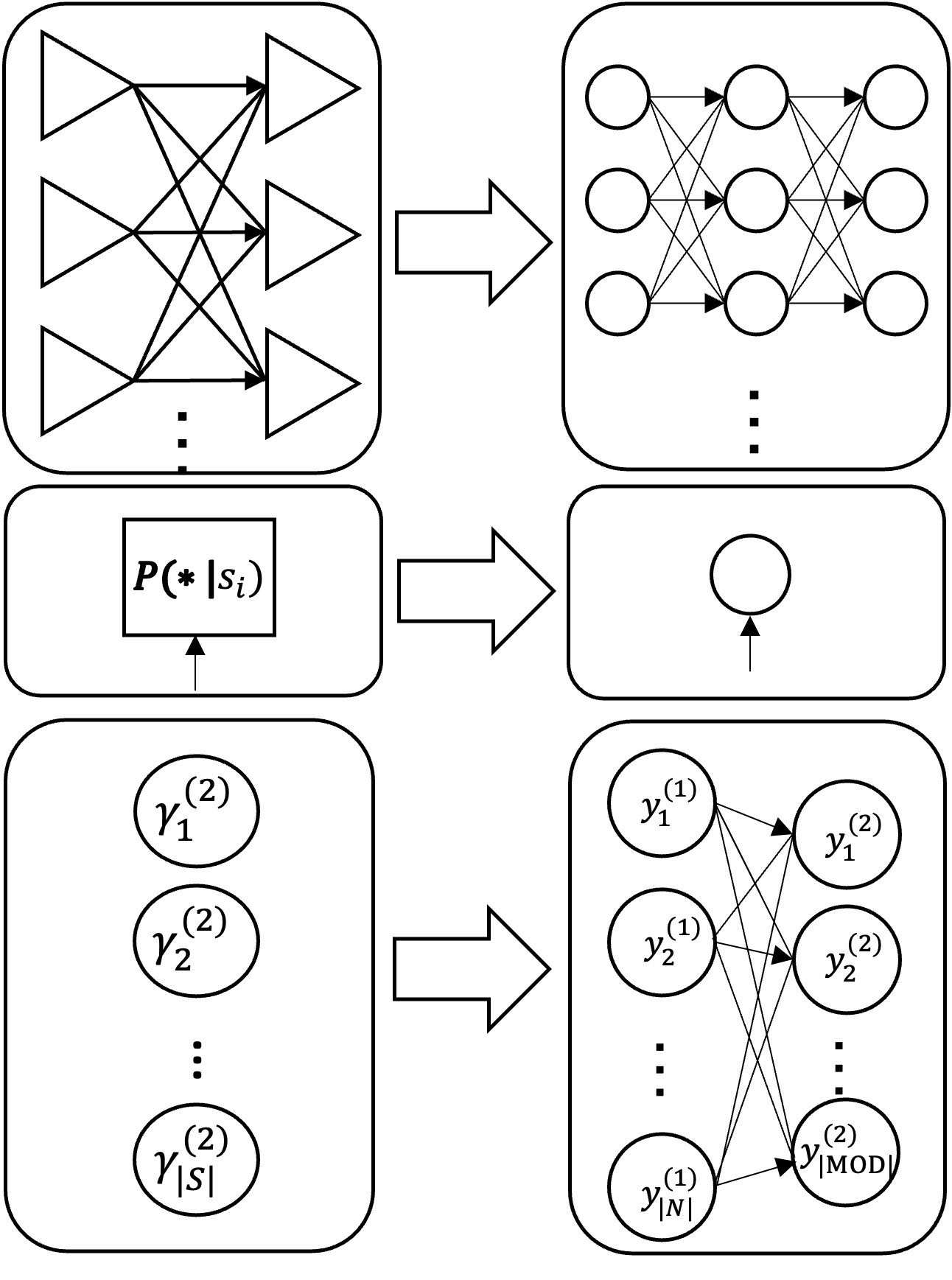}  
  \label{fig:blockChange}
  }
\subfloat[]{
  \includegraphics[width=0.75\textwidth]{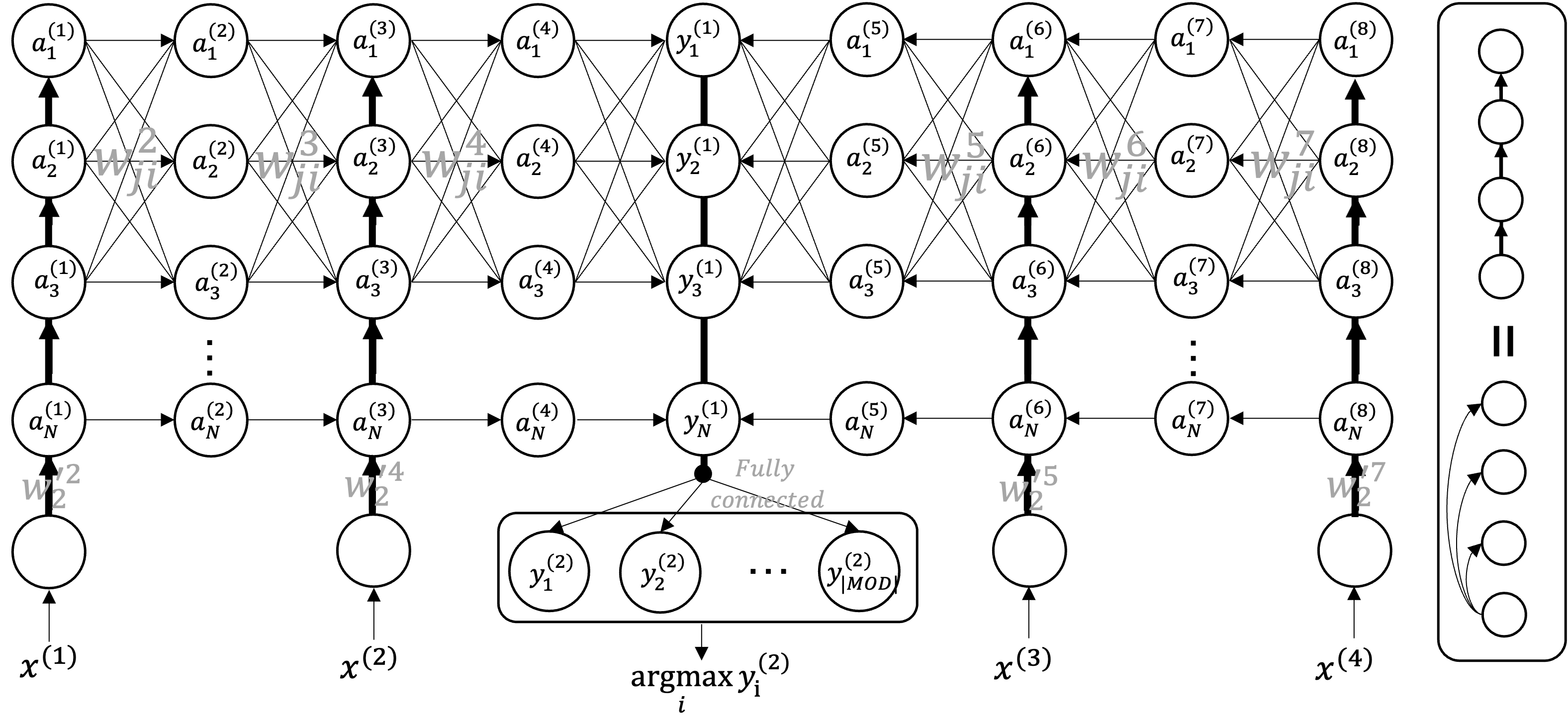}  
  \label{fig:blockNeq}
}
\caption{(a) Mapping FB to neural network: i) products required to compute $\alpha$'s and $\beta$'s are replaced with two-layer FCLs, ii) computing conditional probability is replaced with a single perceptron, and iii) products to compute $\gamma$'s are replaced with two-layer FCLs. (b) Proposed neural network based equalizer, NeuralEQ. (Note: thick lines are used to represent full connections between components.)}
\label{fig:blockdiagrams}
\end{figure*}

\subsection{Architecture Description}

The proposed NeuralEQ is designed to have a similar structure to the FB algorithm. 
Fig.~\ref{fig:blockFwdBwd} shows an illustration of the FB algorithm computation, 
assuming that one symbol is decoded from four received values. 
As expressed in \eqref{equa:fwdbwd}, 
$\alpha_i^{t}$ is composed of the multiplication and 
addition of $\alpha^{t-1}_j$, $a_{ij}$, and $P(X^t|S^t)$, 
which are expressed by the arrow lines between the triangles in Fig.~\ref{fig:blockFwdBwd}. 
$\beta_i^t$ calculated in the backward path is also drawn in the same way, 
and only the direction is opposite. 
Finally, the part where $\alpha$'s and $\beta$'s are multiplied to obtain $\gamma$'s and to find the state with the highest probability is shown in the middle of Fig.~\ref{fig:blockFwdBwd}.


To replace each component in Fig.~\ref{fig:blockFwdBwd} with a neural network, 
we first note that, in \eqref{equa:fwdbwd} (i.e. $\alpha _{i}^{t} = \sum _{j\in |\mathcal{S}|} \alpha _{j}^{t-1}a_{ji}P(x^{t}|S^{t}=s_j)$), 
the terms, $\alpha_j^{t-1}$, $a_{ji}$, and $P(x^t|s_j)$, are multiplied together, 
where $P(x^t|s_j)$ and $\alpha_j^{t-1}$ are functions of observed sequence $x^{1:t}$, and $a_{ji}$'s are constant given the system. 
In the proposed neural network, this multiplication is replaced with FCLs with two-layer depth, 
and calculating $P(x^t|s_j)$ is replaced with a single perceptron.
Finally, we replace \eqref{equa:gamma} with two FCLs whose output layer has the size of the number of transmitted modulation levels, $|MOD|$. 
For example, if PAM4 is used, then the output layer has four neurons.
These replacements are described in Fig.~\ref{fig:blockChange}.
Note that the number of neurons per layer, which is expressed as $N$ in Fig.~\ref{fig:blockNeq}, is not the same as the total number of hidden states $|\mathcal{S}|$ in the proposed architecture. 
Smaller $N$ is desirable to reduce the amount of computation, which will be described in more detail in Section~\ref{sec:complextiy}.

\subsection{Comparison with FCL-based Architecture} \label{subsec:comp}

Compared to RNN-based equalizers, both NeuralEQ and FCL-based equalizer have no feedback structure, 
and they employ only feed-forward architecture (see Fig.~\ref{fig:blockNeq}), 
which allows efficient pipelining for high-speed implementation.
Although NeuralEQ and FCL-based equalizer can be both implemented for high-speed wireline communication,
NeuralEQ has more compact and energy-efficient architecture than FCL-based ones. 
As shown in Fig.~\ref{fig:blockFwdBwd}, when the FB algorithm decodes the target symbol given a sequence of channel output, 
the output values further away from the target need to go through deeper calculations compared to the closer ones.
However, in the case of FCLs, all the channel outputs (i.e. the network inputs) will have a uniform computational depth. 
From the observations from the FB algorithm, which is an optimal symbol decoder, it can be inferred that if FCL-based architecture is employed for equalization, 
parameters would be wasted for the network inputs close to the target symbol.
To address this issue, NeuralEQ reflects this imbalance in computational depths, having more efficient architecture for high-speed wireline communication, where computational complexity is an essential factor to consider. 

The proposed NeuralEQ has a hyperparameter $N$, which is the number of neurons per layer.
Choosing an optimal $N$ is important since the computational complexity of NeuralEQ is heavily dependent on $N$. 
There are many ongoing studies about optimizing hyperparameters of the network, and researchers have provided different types of search strategies depending on the network architecture. 
We choose Tree-Structured Parzan Estimator (TPE)~\cite{bergstra2011algorithms} to optimize $N$ because it is an appropriate algorithm for a model which has low dimensional search space, and NeuralEQ has a single hyperparameter $N$.

For performance and complexity comparison, we choose an FCL-based equalizer with two hidden layers as a baseline. 
Hyperparameters (i.e. number of neurons in each layer) of the FCL-based equalizer are also found using TPE~\cite{bergstra2011algorithms} by which BER can be minimized.
NeuralEQ and FCL-based one are tested with four different channel loss, ranging from 7\,dB to 21\,dB at the Nyquist frequency. 
In Table~\ref{tab:tpe}, for each channel loss, the tested hyperparmeters of NeuralEQ and FCL and their BER performance are tabulated.
Table.~\ref{tab:tpe} clearly shows that, for all four channel conditions, the proposed NeuralEQ has much lower number of parameters, or computational complexity, than the FCL-based equalizer. 
Note that the number of parameters in NeuralEQ is only 7.5\,\% to 27.0\,\%,
it shows even better BER performance over FCL-based equalizers for all the channels,
which implies that the proposed architecture of NeuralEQ is computationally efficient compared to FCL-based equalizers.

\ignore{
Compared to RNN-based equalizers, the proposed NeuralEQ has no feedback structure, 
which is a critical design overhead for high-speed operation. 
Because the feedback delay must meet 1UI timing constraint, it usually incurs huge area and power overhead. 
For this reason, recent papers~\cite{krupnik2020112,im2020112,yoo20206,li2023100,xu20218} adopt only 1-tap or 2-tap DFE despite of its performance.
On the other hand, the proposed neural network is only consists of feed-forward path as shown in Fig.~\ref{fig:blockNeq}, 
so efficient pipelining in implementation is possible.
}

\begin{table*}[]
\begin{center}
\caption{Complexity and performance comparison between NeuralEQ and FCL-based equalizer.} 
\label{tab:tpe}
\resizebox{0.9\textwidth}{!}{
\begin{tabular}{lllllll}
\hline 
\multicolumn{1}{c}{\bf Channel loss (dB)} &\multicolumn{3}{c}{\bf FCL-based EQ} &\multicolumn{3}{c}{\bf NeuralEQ}
\\
\multicolumn{1}{c}{\bf w/ SNR=11\,dB} &\multicolumn{1}{c}{\bf Hyper-param} &\multicolumn{1}{c}{\bf BER (e-3)} &\multicolumn{1}{c}{\bf \# of param} &\multicolumn{1}{c}{\bf Hyper-param} &\multicolumn{1}{c}{\bf BER (e-3)} &\multicolumn{1}{c}{\bf \# of param}
\\
\hline
\multicolumn{1}{c}{7} &\multicolumn{1}{c}{[ 216, 376 ]} &\multicolumn{1}{c}{1.126} &\multicolumn{1}{c}{85,908} &\multicolumn{1}{c}{N: 22} &\multicolumn{1}{c}{0.961} &\multicolumn{1}{c}{18,594 (21.6\%)}
\\
\multicolumn{1}{c}{12} &\multicolumn{1}{c}{[ 408, 488 ]} &\multicolumn{1}{c}{2.845} &\multicolumn{1}{c}{206,852} &\multicolumn{1}{c}{N: 20} &\multicolumn{1}{c}{2.309} &\multicolumn{1}{c}{15,464 (7.47\%)}
\\
\multicolumn{1}{c}{16} &\multicolumn{1}{c}{[ 472, 344 ]} &\multicolumn{1}{c}{12.56} &\multicolumn{1}{c}{170,228} &\multicolumn{1}{c}{N: 35} &\multicolumn{1}{c}{11.221} &\multicolumn{1}{c}{45,959 (27.0\%)}
\\
\multicolumn{1}{c}{21} &\multicolumn{1}{c}{[ 352, 512 ]} &\multicolumn{1}{c}{56.054} &\multicolumn{1}{c}{187,364} &\multicolumn{1}{c}{N: 29} &\multicolumn{1}{c}{50.516} &\multicolumn{1}{c}{31.817 (16.9\%)}
\\
\hline
\end{tabular}
}
\end{center}
\end{table*}

\subsection{Choosing NeuralEQ Input Size and Decoded Symbol Position}
\label{sec:choice}

In principle, the FB algorithm can decode the symbol at any position from the input sequence. 
Even though Fig.~\ref{fig:blockFwdBwd} describes the case when decoding the second symbol, 
it can calculate $\gamma^t$'s at any time $t$, $1\leq t \leq 4$. 
However, the decoding error probability will not be uniform for all positions, 
e.g. the error probability of decoding the last symbol in Fig.~\ref{fig:blockFwdBwd} will be generally much higher than that of decoding the second or third symbol.
This is because wireline channels typically have both pre-cursors and post-cursors around the main-cursor, 
and for a given length of input sequence, 
as the target symbol position moves to the left (or right), 
the information about the pre-cursors (or post-cursors) is lost. 
In other words, the position of the target symbol must be carefully chosen for good BER performance.

Let the channel input sequence $\mathbf{z}$, the observed noisy channel output $\mathbf{x}$, the channel impulse response $\mathbf{h}$, and the additive white noise $\mathbf{n}\sim \mathcal{N}(0,\sigma ^{2}\mathbf{I})$,
i.e. $\mathbf{x} = \mathbf{z}*\mathbf{h}+\mathbf{n}$. 
If we denote the NeuralEQ input size as $T$ and the target symbol position as $D$, 
NeuralEQ estimates $z^{t+D}$ for the given input sequence $x^{t:t+T-1}$,
i.e. $\hat{z}^{t+D} = f_{\mathbf{NEQ}}(x^{t:t+T-1})$, 
where $f_{\mathbf{NEQ}}$ indicates the function represented by NeuralEQ.
If there are pre-cursors in the channel impulse response $\mathbf{h}$, 
in order to account for the delay between $\mathbf{z}$ and $\mathbf{x}$, 
NeuralEQ estimates the symbol at the following:
\begin{equation}
    \hat{z}^{t+D-|h_{pre}|} = f_{\mathbf{NEQ}}(x^{t:t+T-1})
    \label{eq:infer}
\end{equation}
where $|h_{pre}|$ is the number of pre-cursors in ISI.
In this paper, we choose the length of the input sequence $T$ and the target symbol position $D$ as 12 and 4, respectively. 
This generally shows good performance over various wireline channels tested. 
For the channels with much larger/less ISI compared to the tested ones, 
the input size and target symbol position should be adjusted.

\ignore{
Let the channel input sequence $\mathbf{z}$, the observed noisy channel output $\mathbf{x}$, the channel impulse response $\mathbf{h}$, and the additive white noise $\mathbf{n}\sim \mathcal{N}(0,\sigma ^{2}\mathbf{I})$,
i.e. $\mathbf{x} = \mathbf{z}*\mathbf{h}+\mathbf{n}$. 
If we denote the NeuralEQ input size as $T$ and the target symbol position as $D$, the estimator $\hat{\mathbf{z}}$ for the original signal $\mathbf{z}$ can be expressed as $\hat{z}^{t+D} = f_{\mathbf{w}}(x^{t:t+T-1})$. Note that $f_{\mathbf{w}}$ indicates the function of the proposed neural network, NeuralEQ.
However, if there are pre-cursors in the channel ISI, 
delay occurs between $\hat{\mathbf{z}}$ and $\hat{\mathbf{x}}$, 
the final equation is summarized again as follows.
\begin{equation}
    \hat{z}^{t+D-|pre\_cursors|} = f_{\mathbf{w}}(x^{t:t+T-1})
    \label{eq:infer}
\end{equation}
where $|pre\_cursors|$ is the number of pre-cursors in ISI.
}

\begin{table*}[t]
\centering
\caption{Comparison of the number of computations between FB and NeuralEQ.\label{tab:complexity}}
\resizebox{0.9\textwidth}{!}{
\begin{tabular}{lllll}
\hline 
\multicolumn{1}{c}{\bf } &\multicolumn{2}{c}{\bf FB algorithm}  &\multicolumn{2}{c}{\bf NeuralEQ}
\\ \hline 
\makecell[c]{forward-path equation}
&\multicolumn{2}{l}{$\alpha _{i}^{t} = \sum _{j\in |S|} \alpha _{j}^{t-1}a_{ji}P(x^{t}|s_{i})$}   
&\multicolumn{2}{l}{\makecell[l]{
$a_{i}^{t} = \tanh(\sum _{j\in N}{ a_{j}^{t-1}{w_{ji}}^t}+Mw_{N+1,i}+b_{i}^t)$\\
$a_{i}^{t+1} = \tanh(\sum _{j\in N}{ a_{j}^{t}w_{ji}^{t+1} + b_{i}^{t+1}})$\\
where $M=\tanh({{w'}_{i}^{t}}x^{t} + {b'}_{i}^{t})w_{i}^t )$\\
}
}
\\ \hline 
\# of multipliers
&\makecell[l]{$|S|^{2}+|S|$}
&\makecell[l]{$1.10\times10^{12}$} 
&\makecell[l]{$2N^{2}+N$}
&\makecell[l]{$2.08\times10^{3}$} 
\\
\# of adders           
&\makecell[l]{$|S|^{2}-|S|$}
&\makecell[l]{$1.10\times10^{12}$} 
&\makecell[l]{$2N^{2}+N$}
&\makecell[l]{$2.08\times10^{3}$} 
\\
\# of conditional prob.
&\makecell[l]{$|S|^{2}$}
&\makecell[l]{$1.10\times10^{12}$}
&\makecell[l]{N/A}
&\makecell[l]{N/A}
\\
\# of tanhs
&\makecell[l]{N/A}
&\makecell[l]{N/A}
&\makecell[l]{$3N$}
&\makecell[l]{$9.6\times10^{1}$} 
\\
\hline
\end{tabular}
}
\end{table*}

\subsection{Complexity Comparison with FB Algorithm}
\label{sec:complextiy}

One of the main motivations of this work is to find a neural network architecture that mimics the FB algorithm, while having less computational complexity. 
To analyze the computational complexity of NeuralEQ, we only compare the forward path equation of the FB algorithm and NeuralEQ. 
Although both the FB algorithm and NeuralEQ consist of the forward path, backward path, and $\gamma$ computation, forward and backward paths have identical structures and computing $\gamma$'s is negligible. 

The amount of computation for FB and Neural is summarized in Table~\ref{tab:complexity}. 
Specifically, the operator type and number of operations required to calculate the forward-path recursion (i.e. computing $\alpha_i^t$ from $\alpha_j^{t-1}(j\in |\mathcal{S}|)$ in FB and computing $a _i^{t+1}$ from $a_j^{t-1}(1 \leq j \leq N)$ are shown in Table~\ref{tab:complexity}.
Notations are referred to Fig.~\ref{fig:blockdiagrams}. 
Note that, because one layer of the FB algorithm is replaced with two layers of FCL (see Fig.~\ref{fig:blockChange}), the forward-path equation of NeuralEQ includes the two-layer computation, i.e. $a^{t-1}$ to $a^{t+1}$.
The required number of multiplication in the FB algorithm, which is a dominant factor in computation complexity, is $|\mathcal{S}|^2+|\mathcal{S}|$. 
On the other hand, NeuralEQ requires $2N^2+N$ multiplications.

If $N$ and $|\mathcal{S}|$ are the same, NeuralEQ has more multiplication, but in practice $N$ is much smaller than $|\mathcal{S}|$. 
The numerical values in Table~\ref{tab:complexity} represent the number of each operator for the case when PAM4 modulation is used, $|\mathbf{h}|=10$ (i.e. the total number of pre-, main-, and post-cursors are 10), and $N$ = 32.
In this case, since $|\mathcal{S}|$ is determined by $|MOD|^{|\mathbf{h}|}$, $|\mathcal{S}|=4^{10}=2^{20}$. 
However, with the same channel, we found that NeuralEQ with $N=32$ has comparable BER performance. 
Thus, the number of multiplication is $2^{11}+32$, which means NeuralEQ has around $2^{9}$ times fewer multiplication compared to the FB algorithm.
As shown in Table~\ref{tab:complexity}, NeuralEQ requires much lower number of other operators as well as multiplication.
Moreover, we will discuss a method to further reduce complexity in Section~\ref{sec:prune}.

\subsection{Training NeuralEQ}
NeuralEQ is trained using the following loss function,
\begin{equation}
    \mathrm{loss} = \frac{1}{N}\sum_{t=1}^{N} \mathrm{CE}(f_{\mathbf{NEQ}}(x^{t:t+T-1}), z^{t+D-|h_{pre}|})
\end{equation}
where $N$ is the batchsize, CE represents the cross-entropy loss, and we use the same notations introduced in Section~\ref{sec:choice}. 
In general, the trained network may not perform well due to the limited number of training data. 
We found that NeuralEQ also has a similar issue. 
Therefore, securing a large number of training set $\{x^{t:t+T-1}, z^{t+D-|h_{pre}|}\}$ is advantageous for training. 
The good news is that in wireline communication, 
as long as data is continuously transmitted from the transmitter with known pattern, 
it is possible to generate training data as much as we want. 
Because training data could be infinitely generated, 
there was no reason to repeatedly use the same training data. 
Thus, in NeuralEQ training, there is no notion of ‘epoch’, or rather, the network is trained in a single epoch. 
Every time when training data is needed, it is freshly generated.
Training, validation, and test sets were each randomly generated with the same channel impulse response and SNR. 
Random data sequence of length 2e9, 2e8, and 1e7 were used for training, validation, and test, respectively. 
The training parameters are summarized in Table~\ref{tab:params}.

\begin{figure}[t]
\begin{center}
\centerline{\includegraphics[width=0.9\columnwidth]{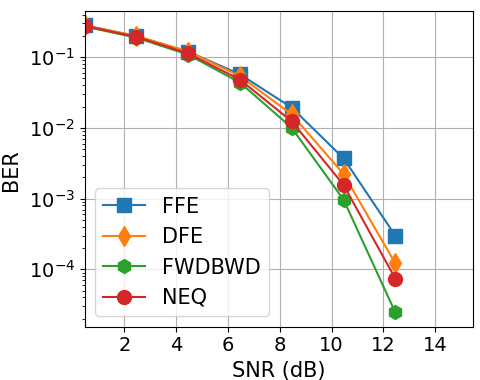}}
\caption{Performance comparison between FFE, DFE, FB, and NeuralEQ.}
\label{fig:isi0}
\end{center}
\end{figure}

To examine the performance of the proposed NeuralEQ, 
BER vs SNR curve of the trained NeuralEQ is obtained under the same environment as in Fig.~\ref{fig:isi0_woNEQ}, i.e. $h[n]=[1.0,0.4,0.2,0.1]$.
Fig.~\ref{fig:isi0} shows the comparison between the 8-tap FFE, the 3-tap DFE, the FB algorithm, and the trained NeuralEQ.
It is shown that NeuralEQ has superior performance to FFE and DFE
and has the BER performance close to the optimal decoder, the FB algorithm. 
The observation that BER of NeuralEQ is lower than that of conventional equalizers is validated in the following section by testing NeuralEQ over various practical channels.

\begin{table}[t]
\centering
\caption{Proposed NeuralEQ hyperparameters.}
\resizebox{0.8\columnwidth}{!}{
    \begin{tabular}{cc}
     \\ \hline
      \bf{Network parameters} & \bf{Values} \\
      \hline
        Input size & 12 \\
        Target symbol position ($D$) & 4 \\
        \# of neurons per layer ($N$) & 32 \\
        \hline
       \bf{Training parameters} & \bf{Values} \\
       \hline
        \# of train, valid, test data & 2e9, 2e8, 1e7 \\
        Batchsize & 8192 \\
        Loss function & Cross Entropy \\
        Optimizer & Adam \\
        Learning rate & 1e-3 \\ \hline
      \end{tabular}
      \label{tab:params}

}
\end{table}

\section{Simulation results}
\label{sec:results}
\subsection{Performance on Various Channels}

\begin{figure}[t]
\centering
\subfloat[]{
\includegraphics[width=0.24\textwidth]{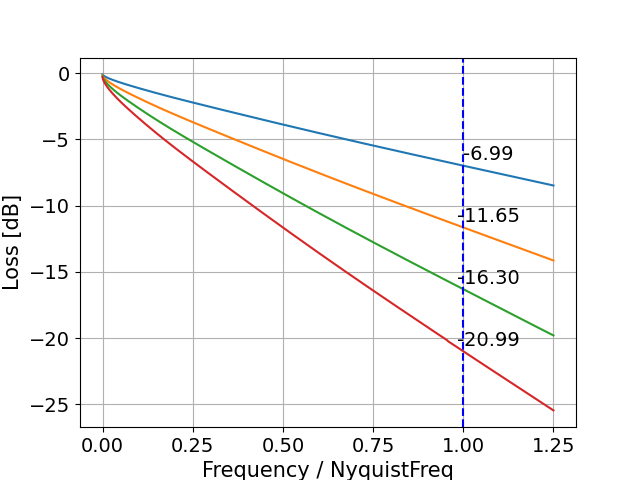}
\label{fig:chfreq}
}
\subfloat[]{
\includegraphics[width=0.24\textwidth]{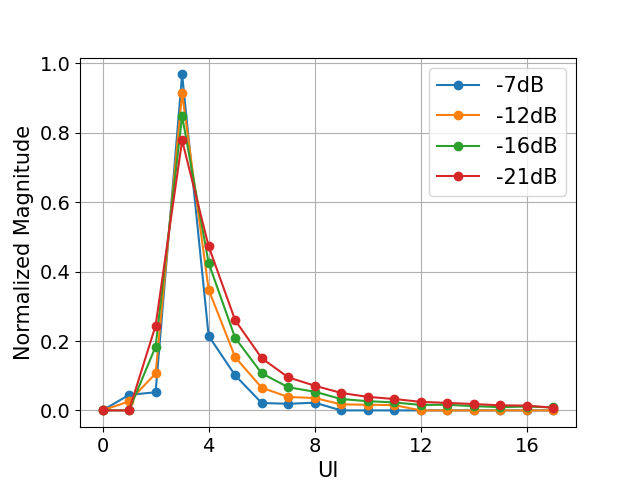}
\label{fig:chsbr}
}
\caption{(a) Frequency response, and (b) discrete-time impulse response of tested channels.}
\label{fig:ch}
\end{figure}

Four different channel characteristics were extracted from real PCB striplines to evaluate the performance of the proposed NeuralEQ. 
Each channel has -7\,dB, -12\,dB, -16\,dB, and -21\,dB loss at the Nyquist frequency, respectively (see Fig.~\ref{fig:chfreq}), 
and each channel has a discrete-time impulse response as shown in Fig.~\ref{fig:chsbr}.
Unlike the channel used in Section~\ref{sec:proposed}, tested real channels in this section have pre-cursors, so using FFE or DFE alone for equalization does not show good enough performance. 
Hence, for comparison, performance of an equalizer employing both FFE and DFE, which is a typical choice for high-speed wireline communication with lossy channels to minimize both pre- and post-cursors, is also evaluated. 
The number of FFE taps is set as 24, and the number of DFE taps is 5, which is more than what is actually used in the PAM4 wireline communication systems~\cite{9366063,9134399,8952650} so that there can be no performance degradation due to lack of taps. 
The FB algorithm is not evaluated for the tested real channels because the number of hidden states for such a long impulse response cannot be handled by a simulation server, causing memory allocation to fail.
The hyperparameter of NeuralEQ are the same as those shown in Table~\ref{tab:params}, except for the channel with -21\,dB loss, which has worse ISI than other channels. 
Thus, in this case, the NeuralEQ input size is increased from 12 to 24.

\begin{figure}[t]
\centering
\subfloat[]{
  \includegraphics[width=0.23\textwidth]{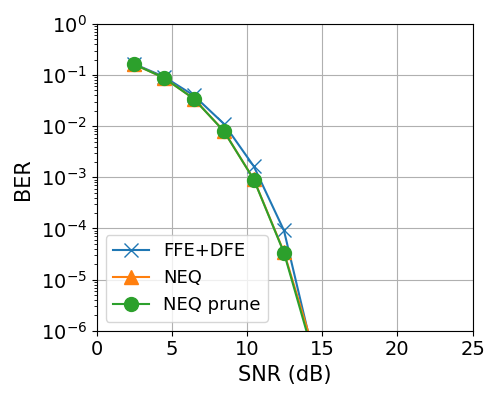}  
  \label{fig:isi7dB_prune_eval}
}
\hspace{-0.5em}
\subfloat[]{
  \includegraphics[width=0.23\textwidth]{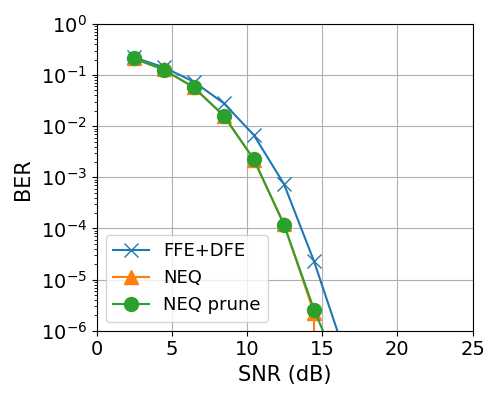}  
  \label{fig:isi12dB_prune_eval}
}
\hspace{-0.5em}
\hfill 
\subfloat[]{
  \includegraphics[width=0.23\textwidth]{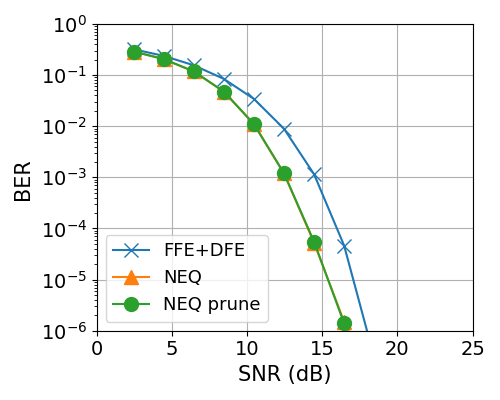}  
  \label{fig:isi16dB_prune_eval}
}
\hspace{-0.5em}
\subfloat[]{
  \includegraphics[width=0.23\textwidth]{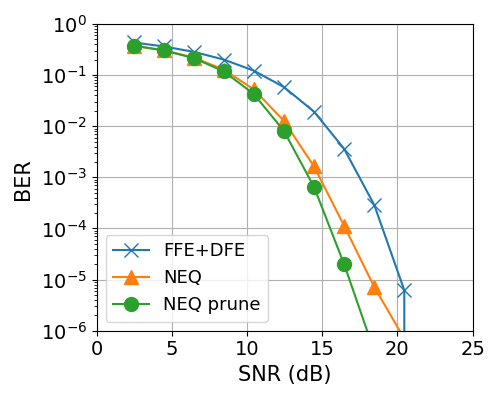}  
  \label{fig:isi21dB_prune_eval}
}
\caption{Performance comparison between FFE+DFE, NeuralEQ, and NeuralEQ with pruning, for channel loss (a) -7\,dB, (b) -12\,dB, (c) -16\,dB, and (d) -21\,dB. In all cases, NeuralEQ has better performance than FFE+DFE with or without pruning. Especially for channel loss -21\,dB, pruning increases the performance of NeuralEQ.}
\label{fig:prune_eval}
\end{figure}

As illustrated in Fig.~\ref{fig:prune_eval}, BER performance of the proposed NeuralEQ is superior to the combination of FFE and DFE with sufficient number of taps in the whole SNR range. 
In particular, the performance gap increases as the channel loss becomes larger, which indicates that NeuralEQ can provide huge performance gain over existing equalizers in recent wireline communication systems that suffer more from high channel loss.
In addition to the four tested channels, we manually generate some channels that has not only ISI but reflection,
and NeuralEQ is found to perform well for such channels also, further highlighting its superior performance compared to conventional equalization techniques such as FFE and DFE.

\subsection{Pruning}
\label{sec:prune}
\ignore{
Although the proposed NeuralEQ needs multiple symbols per single decoding, 
not all input symbols have the same information in decoding the target symbol. 
This is because the magnitude of the cursors decreases as the distance away from the main-cursor, so the information of the target symbol is minor. 
On the other hand, the symbol at the main-cursor position has the most information.
Due to this imbalance in the information of symbol position, even if the complexity of the left and right side layers of the NeuralEQ is lower than those of the middle, it can be expected that NeuralEQ will not significantly degrade performance.
}
The proposed NeuralEQ uses information from multiple symbols for decoding, 
but not all the symbols carry the same level of information. 
This is because the further away a symbol is from the main cursor, the less information it contains about the target symbol. 
Conversely, the symbol at the main cursor position holds the most information. 
This observation implies that performance of NeuralEQ will not be significantly degraded 
even if we reduce the number of parameters in the left and right side layers compared to that in the middle layer.

Pruning~\cite{PhysRevA.39.6600,80236} is an efficient method, which has been widely studied to reduce computational complexity and compress the model by removing weights with small magnitude after training.
In particular, \cite{frankle2018the} has shown that there exist models that have better or similar performance after pruning. 
Therefore, we study whether pruning can be applied to NeuralEQ to reduce the complexity further without degrading performance.

\begin{figure}[t]
    \centering
    \includegraphics[width=\columnwidth]{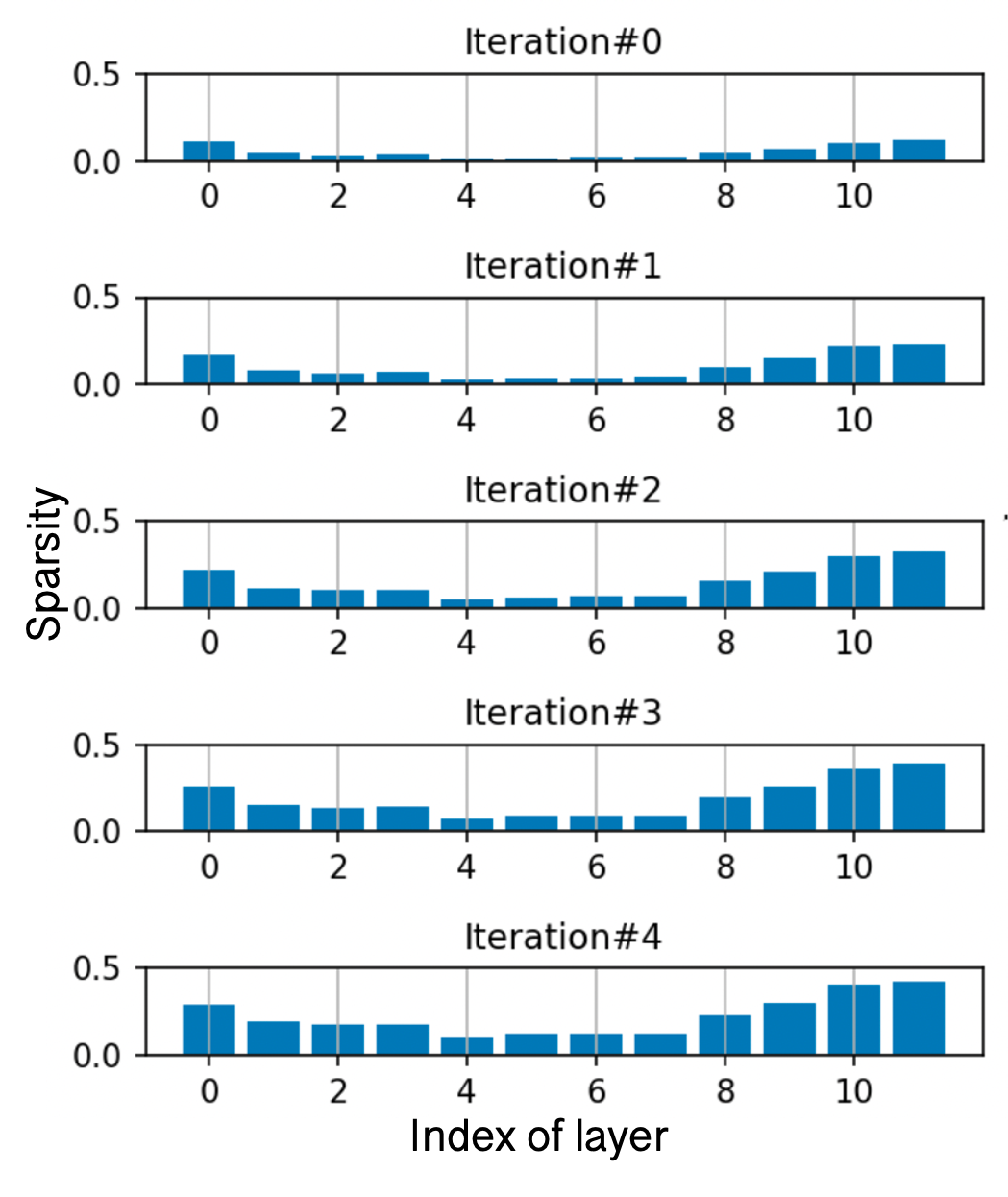}
    \caption{Sparsity level of each layer during pruning iteration. The x-axis indicates the index of layer, and the low index represents the left side layer in Fig.~\ref{fig:blockNeq}, and the high index the right side layer. After each iterations, parameters in the left and right side layers are more pruned than those in the middle.}
    \label{fig:prune_each}
\end{figure}

When pruning is performed on NeuralEQ, small weights or neurons with small weights, which are considered less important, will be removed at each iteration. 
Since the channel impulse response generally has smaller cursor magnitudes as it moves further away from the main-cursor, it can be assumed that the layers at the left and right sides are less important than the middle layer even in NeuralEQ. 
Fig.~\ref{fig:prune_each} shows the sparsity level of each layer during pruning iterations. 
In every pruning iteration, smallest 10\,\% of the total weights are pruned. 
The difference in sparsity level between layers indicates each symbol in the input sequence of NeuralEQ does not contain the same amount of information required for correct decoding.
Pruning removes unimportant weights/neurons especially in the left and right side layers as expected.

\begin{figure}[t]
    \centering
    \includegraphics[width=0.9\columnwidth]{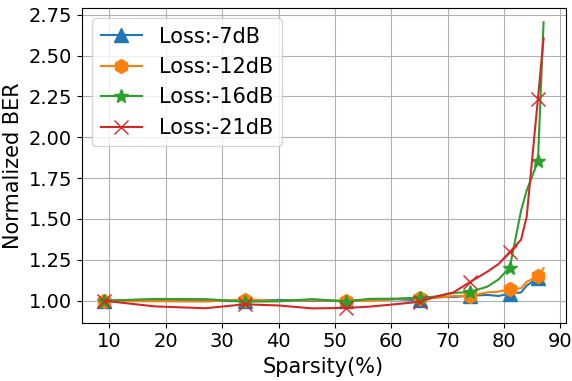}
    \caption{Normalized BER vs sparsity.}
    \label{fig:prune_ber}
\end{figure}

Fig.~\ref{fig:prune_ber} shows how much BER of NeuralEQ is degraded as the sparsity level is increased for the tested four channels.
The Y-axis is the normalized BER (i.e. BER after pruning is divided by BER before pruning), and the X-axis represents the sparsity level. 
Performance degradation due to pruning appears differently for each channel. 
When the channel loss is small, for the -7\,dB-loss channel, performance degradation due to pruning is small, but when the channel loss is high, BER is degraded much once the sparsity level goes over 80\,\%.


\begin{table}[t]
\centering
\caption{Parameter reduction of NeuralEQ after pruning.}
\resizebox{0.9\columnwidth}{!}{
\begin{tabular}{cccc}
\\ \hline 
\makecell{\bf{Channel}\\\bf{loss}} & \makecell{\bf{\# of param.}\\\bf{before prune}} &\makecell{\bf{\# of param.}\\\bf{after prune}} &\makecell{\bf{Diff.}}\\ 
\hline 
-7\,dB & 38,564 & 13,446 & -65\,\% \\ 
-12\,dB &38,564 & 16,600 & -57\,\%  \\ 
-16\,dB &38,564 & 18,445 & -52\,\%  \\ 
-21\,dB & 76,964 & 36,812 & -52\,\% 
\\ \hline
\end{tabular}
}
\label{tab:prune}
\end{table}

In Fig.~\ref{fig:prune_eval}, the BER-vs-SNR curve is also drawn for the pruned NeuralEQ. 
It can be seen that there is no significant difference compared to the performance before pruning and that the pruned NeuralEQ has better performance for the -21\,dB channel. 
The number of parameters before and after pruning is summarized in Table~\ref{tab:prune}. 
From these experiments, pruning is found to be a well-suited methodology to reduce the number of parameters, or complexity, of NeuralEQ without performance degradation.

\begin{figure}[t]
\centering
\subfloat[]{
  \includegraphics[width=0.23\textwidth]{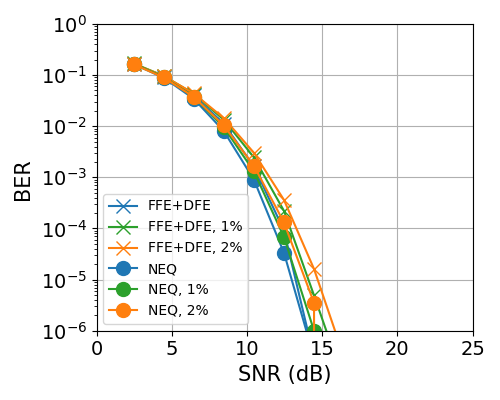}  
  \label{fig:isi7dB_isivari}
}
\hspace{-0.5em}
\subfloat[]{
  \includegraphics[width=0.23\textwidth]{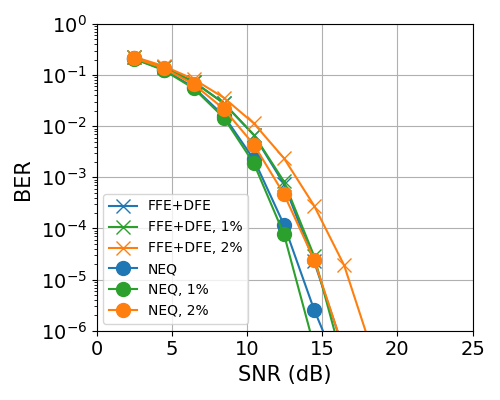}  
  \label{fig:isi12dB_isivari}
}
\hspace{-0.5em}
\subfloat[]{
  \includegraphics[width=0.23\textwidth]{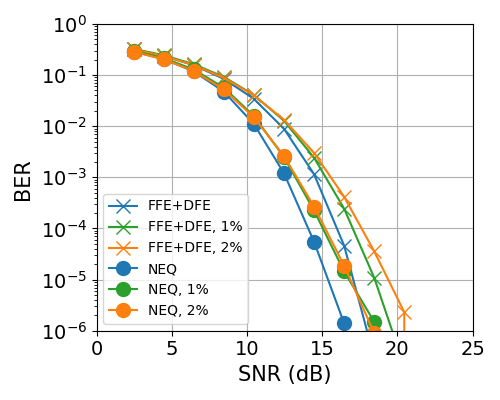}  
  \label{fig:isi16dB_isivari}
}
\hspace{-0.5em}
\subfloat[]{
  \includegraphics[width=0.23\textwidth]{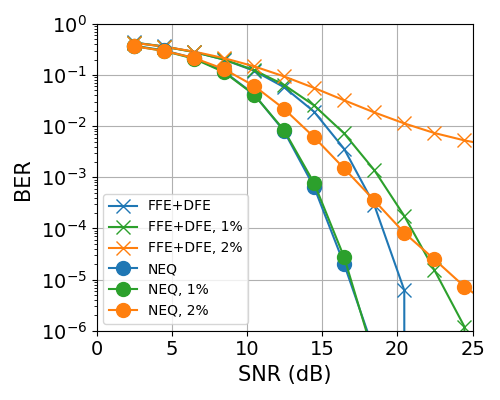}  
  \label{fig:isi21dB_isivari}
}
\caption{Performance variation due to ISI variation for channel loss of (a) -7\,dB, (b) -12\,dB, (c) -16\,dB, and (d) -21\,dB. FFE+DFE and NeuralEQ are evaluated for comparison, and the magnitude of skew is varied with 0\,\%, 1\,\%, and 2\,\%. Regardless of ISI variation, NeuralEQ has better performance than conventional equalizers.  }
\label{fig:isivari}
\end{figure}

\ignore{  
\begin{figure*}[h]
\begin{subfigure}{.25\textwidth}
  \centering
  \includegraphics[width=\textwidth]{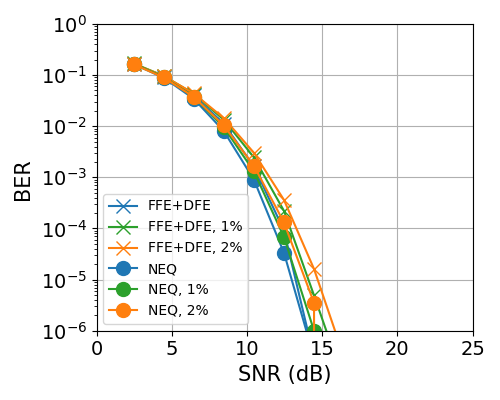}  
  \caption{}
  \label{fig:isi7dB_isivari}
\end{subfigure}
\hspace{-0.5em}
\begin{subfigure}{.25\textwidth}
  \centering
  \includegraphics[width=\textwidth]{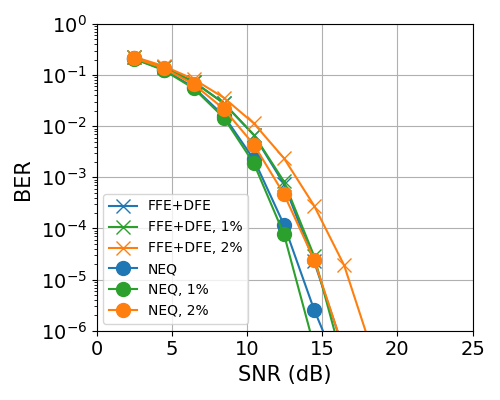}  
  \caption{}
  \label{fig:isi12dB_isivari}
\end{subfigure}
\hspace{-0.5em}
\begin{subfigure}{.25\textwidth}
  \centering
  \includegraphics[width=\textwidth]{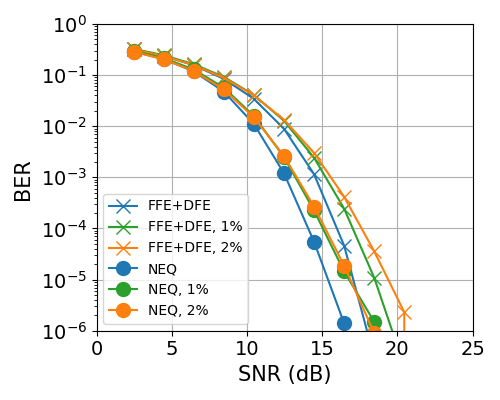}  
  \caption{}
  \label{fig:isi16dB_isivari}
\end{subfigure}
\hspace{-0.5em}
\begin{subfigure}{.25\textwidth}
  \centering
  \includegraphics[width=\textwidth]{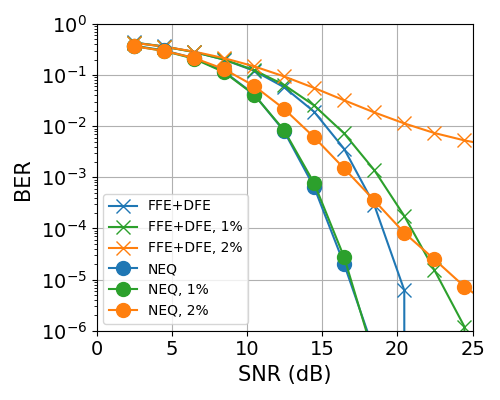}  
  \caption{}
  \label{fig:isi21dB_isivari}
\end{subfigure}
\caption{Performance variation due to ISI skew for channel loss (a) -7dB, (b) -12dB, (c) -16dB, and (d) -21dB, which occurs in practical application. Both conventional EQ and NeuralEQ are evaluated for comparison, and the magnitude of skew of ISI is varied with 0\%, 1\%, and 2\%. Regardless of ISI skew, NeuralEQ has better performance than conventional EQ.  }
\label{fig:isivari}
\end{figure*}
}

\subsection{Robustness to ISI variation}

In practice, the characteristics of the channel used during training and during inference may differ slightly due to the factors such as environmental (e.g. temperature or humidity) variation, and inherent channel variability. 
Hence, it is necessary to evaluate and ensure the robustness of the proposed NeuralEQ against channel variation. 
In order to quantify this, training and test data are generated independently from two different channels having small difference in the impulse response, 
we study how much BER of NeurlEQ is degraded as the difference in channel impulse response gets larger.
The combination of FFE and DFE is also evaluated in the same environment as a comparison baseline. 

The amount of difference in the channel impulse response is given as follows. 
When the impulse response of the channel is $\mathbf{h}$ and let the difference, or skew, be $\mathbf{s}$, $\mathbf{h_{skew}} = \mathbf{h} + \mathbf{s}$, where $\mathbf{s}\sim \mathcal{N}(0,\sigma ^{2}\mathbf{I})$, and $\sigma=\max ({\mathbf{h}})p\ (p\in \{0,0.01,0.02\})$, which is proportional to the magnitude of main-cursor.

Fig.~\ref{fig:isivari} shows the simulation results and comparison between the conventional equalizers and NeuralEQ. 
In the case of -7\,dB, slight channel variation does not affect performance much, but it can be seen that the amount of degradation increases as the channel loss gets larger. 
However, we can observe that NeuralEQ performance is less sensitive to channel variation compared to the conventional equalizers for all the tested channels.
This indicates that NeuralEQ has better tolerance to channel variation than conventional equalizers.
\section{Conclusion}

This paper presents a neural-network-based equalizer NeuralEQ for high-speed wireline communication systems. 
The proposed NeuralEQ architecture mimics the FB algorithm, but it has much less computational complexity than the FB algorithm and shows better BER performance than existing equalizers such as FFEs and DFEs. 
Unlike DFEs or RNN-based equalizers, NeuralEQ is composed of only feed-forward layers without any feedback structure, which makes it amenable to high-speed implementation.
Moreover, complexity of NeuralEQ can be further reduced using pruning.
To validate the effectiveness of NeuralEQ, performance of conventional equalizers and NeuralEQ is compared over four practical channels with different loss.
NeuralEQ is found to have better BER performance and higher tolerance to channel variation compared to conventional equalizers.

\bibliographystyle{IEEEtran}
\bibliography{ref}

\vfill

\end{document}